\documentclass[%
twocolumn,
showpacs,
nofootinbib,
nobibnotes,
amsmath,amssymb,
aps,
pra,
floatfix,
]{revtex4}
\usepackage{graphicx}
\usepackage{dcolumn}
\usepackage{bm}
\usepackage{hyperref}
\usepackage{natbib}
\usepackage{subfigure}

\begin{document}

\preprint{APS/123-QED}

\title{Transfer of spin squeezing and particle entanglement between  atoms and photons in coupled cavities via two-photon exchange}
\author{Ali \"{U}. C. Hardal}
\author{\"{O}zg\"{u}r E. M\"{u}stecapl{\i}o\u{g}lu}
\email{omustecap@ku.edu.tr}
\affiliation{ Department of Physics, Ko\c{c} University, Sar{\i}yer,
Istanbul, 34450, Turkey}
\date{\today}

\begin{abstract}
We examine transfer of particle entanglement and spin squeezing between atomic and photonic subsystems in optical cavities coupled by two-photon exchange. Each cavity contains a single atom, interacting with cavity photons with a two-photon cascade transition. Particle entanglement is characterized by evaluating optimal spin squeezing inequalities, for the cases of initially separable and entangled two-photon states. It is found that particle entanglement is first generated among the photons in separate cavities and then transferred to the atoms. The underlying mechanism is recognized as an inter-cavity two-axis twisting spin squeezing interaction, induced by two-photon exchange, and its optimal combination with the intra-cavity atom-photon coupling. Relative effect of  non-local two-photon exchange and local atom-photon interactions of cavity photons on the spin squeezing and entanglement transfer is pointed out.
\end{abstract}

\maketitle
\section{Introduction}

Recent investigations of quantum tunneling dynamics of two bosonic particles in photonic  \cite{PhysRevA.83.023814} and atomic \cite{longhi2011optical} settings reveal their rich dynamical features, not available to typical single particle tunneling situation. In the case of ultracold atoms in optical lattices, pairwise tunneling yields rich quantum phase diagrams \cite{PhysRevA.80.013605}. It is proposed that two-boson tunneling can be realized by coupling molecular states to atomic ones. Similar scenario was considered before for spin squeezing \cite{PhysRevA.47.5138,sorensen2001many} and entanglement generation. Entanglement properties of two-photon exchange coupled cavities was analyzed in detail for the cases of two and four dimensional Hilbert space systems. These results could set the pathway towards massively correlated multiphoton nonlinear quantum optical systems \cite{hillery2009introduction,dell2006multiphoton}, which are rapidly developing modern subjects nowadays. The motivation of interest to such systems is their promise in quantum switching, quantum communication and computation and quantum phase transition applications.

Motivated by the link between the two-particle tunneling in atomic systems and spin squeezing, as well as its generation schemes based upon
quantum state transfer between optical and atomic systems \cite{PhysRevA.70.022323,PhysRevLett.78.3221,vernac2001quantum,PhysRevLett.83.1319,PhysRevLett.79.4782,PhysRevA.54.5327},
our aim is to investigate similar spin squeezing route to particle entanglement in two-photon exchange coupled optical cavity system. Previous studies investigate entanglement in the whole atom-photon Hilbert space in two and four dimensions \cite{PhysRevA.83.023814}. Our objective is to examine spin squeezing within photon and atom subsystems and to consider possible transfer of the particle entanglement between the two subsystems.

In this paper, we reveal the inherit two-axis twisting spin squeezing interaction nature of two-photon exchange interaction. The non-local interaction between the cavities induces spin-squeezing in the photon subsystem. Induced spin-squeezing in the photon subsystem is then transferred to the atomic subsystem by the local atom-photon coupling within each cavity. We consider the cases of initially entangled and non-entangled photons. Spin squeezing is witnessed by the optimal spin squeezing inequalities \cite{PhysRevLett.99.250405}.

This paper is organized as follows. In Sec. \ref{sec:spinSqz}, we review the concept of spin squeezing and its relation to particle entanglement. In Sec. \ref{sec:model},
we review the model system introduced in \cite{PhysRevA.83.023814}, and point out that the two-photon hopping interaction Hamiltonian is equivalent to the two-axis twisting Hamiltonian, that can generate squeezed spin states. In Sec. \ref{sec:results}, our results for the spin squeezing and particle entanglement transfer are presented in two subsections for the initially entangled and non-entangled photonic subsystems. The role of quadrature squeezing for these transfers is also pointed out. Finally we conclude in Sec. \ref{sec:conclusion}.

\section{Spin Squeezing and Particle entanglement} \label{sec:spinSqz}
Spin squeezing \cite{ma2011quantum} can be defined in terms of Heisenberg uncertainty. Such a definition is subjective to
choice of coordinate system. Alternative definition, that takes into account quantum correlations between individual atomic spins, is given by Kitagawa and Ueda \cite{PhysRevA.47.5138} in the form
\begin{equation}
\xi=\frac{\Delta J_{\bot}}{\sqrt{J/2}}
\end{equation}
where $J_{\bot}$ is the angular momentum component in the direction of the unit vector $\vec{n}$ along which $\Delta \vec{n}\cdot\vec{J}$ is minimized and $J/2$ is the total variance of each individual spin-1/2. If $\xi<1$ the system is said to be spin squeezed.

Another spin squeezing criteria, which combines spin squeezing with particle entanglement, is given by \cite{sorensen2001many}
\begin{equation}\label{eq:sor}
\xi_{e}^2=\frac{N(\Delta J_{n_{1}})^2}{\langle J_{n_{2}}\rangle^2+\langle J_{n_{3}}\rangle^2}
\end{equation}
where $N$ is the number of particles in the system, $J_{n}=\vec{n}\cdot\vec{J}$ and the $\vec{n}$s are mutually orthogonal unit vectors. The condition $\xi_{e}^2<1$ witnesses not only spin squeezing but also the particle entanglement. Particle entanglement happens in the first quantization and hence is fundamentally different than so called mode entanglement that occurs in second quantization \cite{cunha2007entanglement,van2005single,benatti2011entanglement}.

On the other hand, there exist states for which spin squeezing parameter given in Ineq. \ref{eq:sor} cannot be used to analyze quantum entanglement \cite{oztop2009quantum}. Alternatively, one may use the following inequalities since it has been shown that \cite{PhysRevLett.99.250405} violation of any of them implies entanglement:
\begin{subequations}\label{grp}\label{eq:osis}
\begin{eqnarray}
\langle J_{x}^2\rangle+\langle J_{y}^2\rangle+\langle J_{z}^2\rangle&\leq&\frac{N(N+2)}{4}\\
\label{eq:osi2}(\Delta J_{x})^2+(\Delta J_{y})^2+(\Delta J_{z})^2&\geq&\frac{N}{2}\\
\label{eq:osi3}\langle J_{k}^2\rangle+\langle J_{l}^2\rangle-\frac{N}{2}&\leq&(N-1)(\Delta J_{m})^2\\
\label{eq:osi4}(N-1)[(\Delta J_{k})^2+(\Delta J_{l})^2]&\geq&\langle J_{m}^2\rangle+\frac{N(N-2)}{4}
\end{eqnarray}
\end{subequations}\label{grp}
where $i,j,k$ take all the possible permutations of $x,y,z$. These inequalities are called optimal spin squeezing inequalities. The Ineq. \ref{eq:osi3} is equivalent to the Ineq. \ref{eq:sor} for a system of many particles which has a maximal spin
in some direction \cite{toth2009spin} and they can be related to the positivity of concurrence \cite{PhysRevA.73.062318}. Very recently, these inequalities generalized for qudits with arbitrary spin \cite{PhysRevLett.107.240502}.

Entanglement is considered as quantum information resource, and hence it is desirable to be able transfer entanglement between distant nodes in a quantum network \cite{PhysRevA.70.022323,PhysRevLett.78.3221}. In the following section we shall review the two-photon
hopping model and argue that it can serve as an efficient tool for establishing particle entanglement between distant atoms via
transfer of entanglement from photon subsystem to atom subsystem.

\section{The Model} \label{sec:model}
We consider a system of two spatially separated optical cavities, each containing a single atom,
that is interacting with the cavity field by two-photon transitions. The cavities are coupled to each other with two-photon exchange. The Hamiltonian of the system is written by \cite{PhysRevA.83.023814,wu1997effective,wu1996effective,PhysRevA.52.2218}
\begin{equation} \label{eq:model}
H=H^{(1)}-H_{0}^{(1)}+H^{(2)}-H_{0}^{(2)}+\hbar\zeta(a_{1}^{\dagger2}a_{2}^2+a_{2}^{\dagger2}a_{1}^2),
\end{equation}
where
\begin{eqnarray}
H^{(i)}&=&\hbar\omega a_{i}^{\dagger}a_{i}+\hbar\omega(\sigma_{ee}^{(i)}-\sigma_{gg}^{(i)})
+\hbar\mu\sigma_{ee}^{(i)}\\\nonumber
&+&\hbar\eta\sigma_{gg}^{(i)}+
\hbar\lambda(\sigma_{eg}^{(i)}a_{i}^2+\sigma_{ge}^{(i)}a_{i}^{\dagger2}),
\end{eqnarray}
and
\begin{eqnarray}
H_{0}^{(i)}&=&\hbar\omega(a_{i}^{\dagger}a_{i}+\sigma_{ee}^{(i)}-\sigma_{gg}^{(i)}+(E_{g}+E_{e})/2,
\end{eqnarray}
with $E_{g},E_{e}$ being the energies of the ground and exited states, respectively. The last term in the Eq. \ref{eq:model} is the two-photon exchange interaction between the cavities, characterized by the hopping
rate $\zeta$. The parameters $\mu, \eta$ and $\lambda$ are the free energies of the subsystems written in the convention of Ref. \cite{alexanian1998trapping},
$\sigma_{kl}^{(i)}=|k\rangle^{(i)(i)}\langle l|$ are the atomic transition operators, $k$ and $l$ denotes either ground ($g$) or exited ($e$) states of the atom and $i=1,2$ labels the cavities. Throughout this paper, the hopping rate $\zeta$ will be scaled by the free energy $\lambda$ in numerical calculations so that in all figures where $\zeta\leq1$ corresponds to the local interaction dominant regimes, while $\zeta>1$ corresponds to the hopping dominant regimes.

The spin components for a system of two level atoms defined by \cite{PhysRevLett.79.4782}
\begin{eqnarray}
\nonumber S_{x}&=&\frac{1}{2}(\sigma_{eg}+\sigma_{ge})\\
S_{y}&=&\frac{-i}{2}(\sigma_{eg}-\sigma_{ge})\\
\nonumber S_{z}&=&\frac{1}{2}(\sigma_{ee}-\sigma_{gg}).
\end{eqnarray}
Since we are interested in two coupled cavities each has one atom, we define the spin components of the atoms for this configuration as follows
\begin{eqnarray}
\nonumber S_{x}&=&S_{x}^{(1)}\otimes1^{(2)}+1^{(1)}\otimes S_{x}^{(2)}\\
S_{y}&=&S_{y}^{(1)}\otimes1^{(2)}+1^{(1)}\otimes S_{y}^{(2)}\\
\nonumber S_{z}&=&S_{z}^{(1)}\otimes1^{(2)}+1^{(1)}\otimes S_{z}^{(2)}
\end{eqnarray}
where the superscripts $(1)$ and $(2)$ are label the first and second cavities, respectively.

Likewise, we can consider pseudo-spin operators for the coupled cavity field such that
\begin{eqnarray}
\nonumber L_{x}&\equiv&\frac{1}{2}(a_{1}^{\dagger}a_{2}+a_{2}^{\dagger}a_{1}),\\
 L_{y}&\equiv&\frac{-i}{2}(a_{1}^{\dagger}a_{2}-a_{2}^{\dagger}a_{1}), \\
\nonumber L_{z}&\equiv&\frac{1}{2}(a_{1}^{\dagger}a_{1}-a_{2}^{\dagger}a_{2}).
\end{eqnarray}
They satisfy the SU(2) spin algebra
$[L_\alpha,L_\beta]=\epsilon^{\alpha\beta\gamma}L_\gamma$. Here
$\alpha,\beta,\gamma\in\{x,y,z\}$ and $\epsilon^{\alpha\beta\gamma}$ is the Levi-Civita density.
In terms of $L_{+}\equiv L_{x}+iL_{y}=a_{1}^{\dagger}a_{2}$ and $L_{-}\equiv L_{x}-iL_{y}=a_{2}^{\dagger}a_{1}$, it is straightforward to rewrite two-photon hopping interaction Hamiltonian $H_{p}=\hbar\zeta(a_{1}^{\dagger2}a_{2}^2+a_{2}^{\dagger2}a_{1}^2)$ as
\begin{equation}\label{eq:two_axis}
H_{p}=\hbar\zeta(L_{+}^2+L_{-}^2)=2\hbar\zeta(L_{x}^2-L_{y}^2).
\end{equation}
This is nothing but the two-axis twisting Hamiltonian which squeezes the spin states by twisting them about the two axes \cite{PhysRevA.47.5138}. Clearly neither spin squeezing nor associated particle entanglement can be achieved via one photon hopping in coupled-cavity systems, as the coupling term would be a mere rotation operation generated by $L_x$, though it would produce mode entanglement.

It should be noted that in collective spin systems, i.e., systems in which all spins mutually interact, Eq. \ref{eq:two_axis} is known as the Lipkin-Meshkov-Glick (LMG) Hamiltonian whose entanglement dynamics and exact properties in the thermodynamical limit have been carefully analyzed in Ref. \cite{PhysRevA.70.062304,PhysRevE.78.021106,PhysRevLett.99.050402}. Alternatively, the two-photon hopping can also be seen as a special case of more general nonlinear quantum optical Karassiov-Klimov models \cite{karassiov1994}. Following the Lie-algebraic approach, two-photon hopping term can be recognized by a Jordan-Schwinger map as the difference of raising and lowering operators of the Higgs algebra \cite{higgs1979}. In the view of background field method in the Fokker-Plank formalism \cite{carmichael1986}, we then expect that fluctuations of (Higgs or deformed) effective photon spin operators drive fluctuations of collective atomic spin operators coupled to two-photon transition operators. Fluctuations of two-photon transition operators can be further linked to the quadrature variances. We would then intuitively expect that reduction of spin noise from photon subsystem could be transferred to the atomic subsystem through reduction of quadrature noise. We shall numerically investigate these intuitive expectations for some initial cases in the following section.
\section{Squeezing and Entanglement Transfer} \label{sec:results}
In this section we investigate the transfer of spin squeezing and particle entanglement between distant cavities. To do that, we shall obtain the optimal spin squeezing inequalities for the atomic and photonic systems separately, analyze their variations in time and compare them with each other.

For a given initial state $|\psi(0)\rangle$, restricted to two-photon manifold,
the time-dependent state vector $|\psi(t)\rangle$ may be written as
\begin{eqnarray}
\nonumber|\psi(t)\rangle&=&A(t)|g,2\rangle^{1}|g,0\rangle^{2}+B(t)|g,0\rangle^{1}|g,2\rangle^{2}\\
&+&C(t)|g,0\rangle^{1}|e,0\rangle^{2}+D(t)|e,0\rangle^{1}|g,0\rangle^{2}.
\end{eqnarray}
The general solution of the time-dependent Schr\"{o}dinger equation $i\hbar\partial_{t}|\psi(t)\rangle=H|\psi(t)\rangle$
in the two-dimensional submanifold of the Hilbert space is \cite{PhysRevA.83.023814}
\begin{eqnarray}
\nonumber|\psi(t)\rangle&=&(A_{1}e^{-i\omega_{1}t}+A_{2}e^{-i\omega_{2}t})|\phi_{1}\rangle+(A_{3}e^{-i\omega_{3}t}+A_{4}\\
\nonumber&&e^{-i\omega_{4}t})|\phi_{2}\rangle+(\alpha_{1}A_{1}e^{-i\omega_{1}t}+\alpha_{2}A_{2}e^{-i\omega_{2}t})|\phi_{3}\rangle\\
&&(\alpha_{3}A_{3}e^{-i\omega_{3}t}+\alpha_{4}A_{4}e^{-i\omega_{4}t})|\phi_{4}\rangle
\end{eqnarray}
where $\alpha_{i}=\sqrt{2}/2(\omega_{i}-0.5)$, $\omega_{i}$s are the eigenfrequencies described explicitly in \cite{PhysRevA.83.023814} as well as $A_{i}$s and
\begin{subequations}\label{grp}\label{eq:14}
\begin{eqnarray}
\label{eq:14a}|\phi_{1}\rangle&=&\frac{1}{\sqrt{2}}(|g,2\rangle^{1}|g,0\rangle^{2}+|g,0\rangle^{1}|g,2\rangle^{2})\\
\label{eq:14b}|\phi_{2}\rangle&=&\frac{1}{\sqrt{2}}(|g,2\rangle^{1}|g,0\rangle^{2}-|g,0\rangle^{1}|g,2\rangle^{2})\\
\label{eq:14c}|\phi_{3}\rangle&=&\frac{1}{\sqrt{2}}(|e,0\rangle^{1}|g,0\rangle^{2}+|g,0\rangle^{1}|e,0\rangle^{2})\\
\label{eq:14d}|\phi_{4}\rangle&=&\frac{1}{\sqrt{2}}(|e,0\rangle^{1}|g,0\rangle^{2}-|g,0\rangle^{1}|e,0\rangle^{2}).
\end{eqnarray}
\end{subequations}\label{grp}

The states in Eq. \ref{eq:14} are mutually orthogonal and entangled. Since, this solution allows for Rabi oscillations, one can transfer entanglement between the symmetric ($|\phi_{1}\rangle$ and $|\phi_{3}\rangle$) and antisymmetric ($|\phi_{2}\rangle$ and $|\phi_{4}\rangle$) states separately \cite{PhysRevA.83.023814}.

\subsection{Initially Entangled Photons}
Let us first start with a state which is initially entangled
\begin{equation}
|\psi(0)\rangle=\frac{1}{\sqrt{2}}(|g,2\rangle^{1}|g,0\rangle^{2}+|g,0\rangle^{1}|g,2\rangle^{2}).
\end{equation}
Then, at time $t$, we have
\begin{eqnarray}
\nonumber|\psi(t)\rangle&=&A_{1}(e^{-i\omega_{1}t}+\alpha_{1}^2e^{-i\omega_{2}t})|\phi_{1}\rangle\\
&+&\alpha_{1}A_{1}(e^{-i\omega_{1}t}-e^{-i\omega_{2}t})|\phi_{3}\rangle
\end{eqnarray}
with $A_{1}+\alpha_{1}^2A_{1}=1$. For atom-photon system the density operator, $\rho_{ap}^{12}(t)$, becomes
\begin{eqnarray}
\nonumber \rho_{ap}^{12}(t)&=&|A|^2|\phi_{1}\rangle\langle\phi_{1}|+AB^{*}|\phi_{1}\rangle\langle\phi_{3}|\\
&+&BA^{*}|\phi_{3}\rangle\langle\phi_{1}|+|B|^2|\phi_{3}\rangle\langle\phi_{3}|
\end{eqnarray}
where
\begin{subequations}\label{grp}\label{eq:22}
\begin{eqnarray}
|A|^2&=&A_{1}^2[1+2\alpha_{1}^2\cos{(\delta_{12} t)}+\alpha_{1}^{4}]\\
AB^{*}&=&\alpha_{1}A_{1}^2[1-\alpha_{1}^2+\alpha_{1}^2e^{i\delta_{12} t}-e^{-i\delta_{12} t}]\\
|B|^2&=&2\alpha_{1}^2A_{1}^2[1-\cos(\delta_{12} t)]
\end{eqnarray}
\end{subequations}\label{grp}
and $\delta_{12}=\omega_{1}-\omega_{2}$. To obtain reduced density operators $\rho_{a}^{12}(t)$ and $\rho_{p}^{12}(t)$ for atomic and photonic systems respectively we take partial traces as
\begin{eqnarray}
\nonumber \rho_{a}^{12}(t)&=&Tr_{p}\rho_{ap}^{12}(t)=|A|^2|g\rangle^{11}\langle g|\otimes|g\rangle^{22}\langle g|\\
\nonumber&+&\frac{|B|^2}{2}(|e\rangle^{11}\langle e|\otimes|g\rangle^{22}\langle g|+|e\rangle^{11}\langle g|\otimes|g\rangle^{22}\langle e|\\
&+&|g\rangle^{11}\langle e|\otimes|e\rangle^{22}\langle g|+|g\rangle^{11}\langle g|\otimes|e\rangle^{22}\langle e|)
\end{eqnarray}
and
\begin{eqnarray}
\nonumber\rho_{p}^{12}(t)&=&Tr_{a}\rho_{ap}^{12}(t)=\frac{|A|^2}{2}(|2,0\rangle\langle 2,0|+|2,0\rangle\langle 0,2|\\
&+&|0,2\rangle\langle 2,0|+|0,2\rangle\langle 0,2|)+|B|^2|0,0\rangle\langle 0,0|.
\end{eqnarray}
%
%
\begin{figure}[t]
\includegraphics[width=8cm]{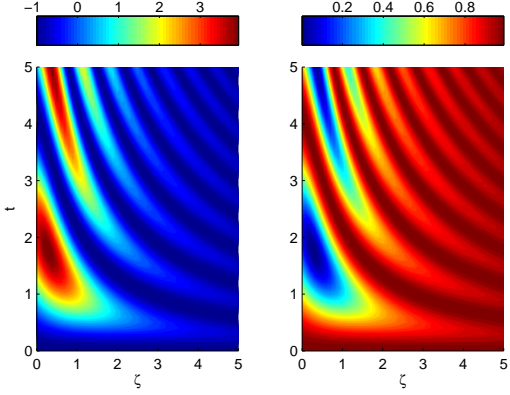}
\caption{(Color online)
Variation of spin squeezing in atomic (left) and photonic (right) subsystems,
with respect to time and hopping constant. $\zeta$ is scaled by $\lambda$ and
$\zeta\leq1$ corresponds to the local interaction dominant regimes, while $\zeta>1$ corresponds to the
hopping dominant regimes. Spin squeezing is characterized
positive values in both figures. At $t=0$, $Ineq_{a}=-1$ and $Ineq_{p}=1$. All the parameters plotted are dimensionless as explained in the text. }
\label{fig:fig1}
\end{figure}
Ineq. \ref{eq:osi4} gives the following inequalities for atomic and photonic subsystems, respectively:
\begin{equation}\label{eq:ineqa}
Ineq_{a}\equiv 4-5|A|^2\leq 0,
\end{equation}
\begin{equation}\label{eq:ineqp}
Ineq_{p}\equiv |A|^2\leq 0,
\end{equation}
where we take $k=x$, $l=z$, $m=y$ in Ineq. \ref{eq:ineqa} and $k=z, l=y$ and $m=x$ in Ineq. \ref{eq:ineqp}. Violation of these inequalities imply spin squeezing (and particle entanglement) in the corresponding systems and as it can be easily seen from the Fig.\ref{fig:fig1}, indeed this is the case. Initially we have non-entangled atoms and entangled photons. As time goes on, atomic system becomes entangled which induced by the photons, i.e., we have entanglement and spin squeezing transfer. Transfers occur periodically in the local-interaction regime ($\zeta<1$). In the hopping dominant regime,
atoms are not entangled or spin squeezed, while the photons are. As should be expected, almost pure two-photon entanglement regimes occur more frequent and in broader intervals with the increase of two-photon hopping rate.
\subsection{Initially Non-Entangled State}
\begin{figure}[t]
\includegraphics[width=8cm]{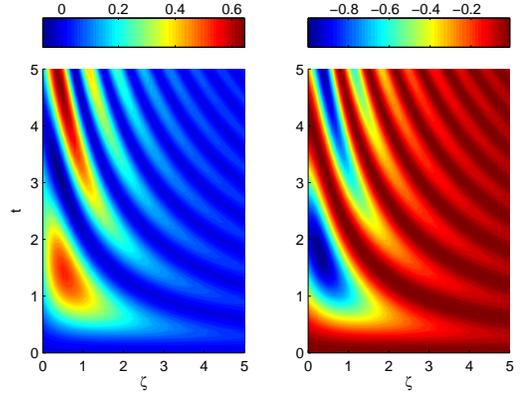}
\caption{(Color Online)
Variations of the spin squeezing in atomic (left) and photonic (right) subsystems with respect to time and hopping constant. Both inequalities are violated, therefore we have both spin squeezed and entangled systems. Spin squeezing is characterized
negative values in both figures. At $t=0$, both inequalities are equal to zero. All the parameters plotted are dimensionless as explained in the text.}
\label{fig:fig2}
\end{figure}

We next consider initially a non-entangled state of the form
\begin{equation}
|\psi(0)\rangle=|g,2\rangle^1|g,0\rangle^2
\end{equation}
which evolves in time as
\begin{eqnarray}
\nonumber|\psi(t)\rangle&=&A_{1}(e^{-i\omega_{1}t}+\alpha_{1}^2e^{-i\omega_{2}t})|\phi_{1}\rangle\\
\nonumber&+&\alpha_{1}A_{1}(e^{-i\omega_{1}t}-e^{-i\omega_{2}t})|\phi_{3}\rangle\\
\nonumber&+&A_{3}(e^{-i\omega_{3}t}+\alpha_{3}^2e^{-i\omega_{4}t})|\phi_{2}\rangle\\
&+&\alpha_{3}A_{3}(e^{-i\omega_{3}t}-e^{-i\omega_{4}t})|\phi_{4}\rangle.
\end{eqnarray}
The density operator, $\rho_{ap}^{12}$, for this case then reads
\begin{widetext}
\begin{eqnarray}
\nonumber\rho_{ap}^{12}(t)&=&|A|^2|\phi_{1}\rangle\langle\phi_{1}|+AB^{*}|\phi_{1}\rangle\langle\phi_{3}|+AC^{*}|\phi_{1}\rangle\langle\phi_{2}|
+AD^{*}|\phi_{1}\rangle\langle\phi_{4}|+BA^{*}|\phi_{3}\rangle\langle\phi_{1}|
+|B|^2|\phi_{3}\rangle\langle\phi_{3}|\\
&+&\nonumber BC^{*}|\phi_{3}\rangle\langle\phi_{2}|+BD^{*}|\phi_{3}\rangle\langle\phi_{4}|
+CA^{*}|\phi_{2}\rangle\langle\phi_{1}|+CB^{*}|\phi_{2}\rangle\langle\phi_{3}|
+|C|^2|\phi_{2}\rangle\langle\phi_{2}|+CD^{*}|\phi_{2}\rangle\langle\phi_{4}|\\
&+&DA^{*}|\phi_{4}\rangle\langle\phi_{1}|+DB^{*}|\phi_{4}\rangle\langle\phi_{3}|+DC^{*}|\phi_{4}\rangle\langle\phi_{2}|+|D|^2\phi_{4}\rangle\langle\phi_{4}|
\end{eqnarray}
\end{widetext}
where
\begin{widetext}
\begin{subequations}\label{grp}
\begin{eqnarray}
AC^{*}&=&A_{1}A_{3}[e^{i(\omega_{3}-\omega_{1})t}+\alpha_{3}^2e^{i(\omega_{4}-\omega_{1})t}+\alpha_{1}^2e^{i(\omega_{3}-\omega_{2})t}+\alpha_{1}^2\alpha_{3}^2e^{i(\omega_{4}-\omega_{2})t}]\\
AD^{*}&=&\alpha_{3}A_{1}A_{3}[e^{i(\omega_{3}-\omega_{1})t}-e^{i(\omega_{4}-\omega_{1})t}+\alpha_{1}^2(e^{i(\omega_{3}-\omega_{2})t}-e^{i(\omega_{4}-\omega_{2})t})]\\
BC^{*}&=&\alpha_{1}A_{1}A_{3}[e^{i(\omega_{3}-\omega_{1})t}-e^{i(\omega_{3}-\omega_{2})t}+\alpha_{3}^2(e^{i(\omega_{4}-\omega_{1})t}-e^{i(\omega_{4}-\omega_{3})t})]\\
BD^{*}&=&\alpha_{1}\alpha_{3}A_{1}A_{3}[e^{i(\omega_{3}-\omega_{1})t}+e^{i(\omega_{4}-\omega_{2})t}-e^{i(\omega_{4}-\omega_{1})t}-e^{i(\omega_{3}-\omega_{2})t}]\\
CD^{*}&=&\alpha_{3}A_{3}^2[1-e^{i(\omega_{4}-\omega_{3})t}+\alpha_{3}^2(e^{i(\omega_{3}-\omega_{4})t}-1)]\\
|C|^2&=&A_{3}^2[1+2\alpha_{3}^2\cos[(\omega_{3}-\omega_{4})t]+\alpha_{3}^4]\\
|D|^2&=&2\alpha_{3}^2A_{3}^2[1-\cos[(\omega_{3}-\omega_{4})t]]
\end{eqnarray}
\end{subequations}\label{grp}
\end{widetext}
with $A_{1}(1+\alpha_{1}^2)=A_{3}(1+\alpha_{3}^2)=1/\sqrt{2}$ and $|A|^2$, $|B|^2$, $AB^{*}$ are given in Eq. \ref{eq:22}. The reduced density operators are given by
\begin{widetext}
\begin{eqnarray}
\nonumber\rho_{a}^{12}(t)&=&(|A|^2+|C|^2)|g\rangle^{11}\langle g|\otimes|g\rangle^{22}\langle g|+\frac{1}{2}\big[(|B|^2+|D|^2+BD^{*}+DB^{*})|e\rangle^{11}\langle e|\otimes|g\rangle^{22}\langle g|\\
\nonumber&+&(|B|^2-|D|^2-BD^{*}+DB^{*})|e\rangle^{11}\langle g|\otimes|g\rangle^{22}\langle e|+(|B|^2-|D|^2+BD^{*}-DB^{*})|g\rangle^{11}\langle e|\otimes|e\rangle^{22}\langle g|\\
&+&(|B|^2+|D|^2-BD^{*}-DB^{*})|g\rangle^{11}\langle g|\otimes|e\rangle^{22}\langle e|\big]
\end{eqnarray}
\begin{eqnarray}
\nonumber\rho_{p}^{12}(t)&=&\frac{1}{2}\big[(|A|^2+AC^{*}+CA^{*})|2,0\rangle\langle2,0|+(|A|^2-AC^{*}+CA^{*})|2,0\rangle\langle0,2|\\
\nonumber&+&(|A|^2+AC^{*}-CA^{*})|0,2\rangle\langle2,0|+(|A|^2-AC^{*}-CA^{*})|0,2\rangle\langle0,2|\big]\\
&+&(2|B|^2+|D|^2)|0,0\rangle\langle0,0|.
\end{eqnarray}
\end{widetext}
%
%
%
%
Then, Ineq. \ref{eq:osi4} gives
\begin{equation}\label{eq:34}
Ineq_{a}\equiv3|B|^2-|D|^2-2(|B|^2+|D|^2)^2\geq0,
\end{equation}
for atomic system and Ineq. \ref{eq:osi2} gives
\begin{equation}\label{eq:35}
Ineq_p\equiv2|A|^2-1\geq0,
\end{equation}
for photonic system, where we take $k=x$, $l=z$ and $m=y$. As in the initially entangled case, violation of these inequalities imply spin squeezing (and particle entanglement) in the corresponding systems. Fig.\ref{fig:fig2} shows that initially unentangled photons quickly get
entangled and this is transferred to the atomic subsystem similar to Fig. \ref{fig:fig1}.
The Fig. \ref{fig:fig2} verifies the periodical nature of the transfers, by anchoring a hopping constant and observing the time evolutions. In terms of particle entanglement, then atomic system becomes more
correlated at the expanse of the correlation of the field particles and vice versa.

Dynamics of correlations can be further interpreted by investigating the quadratures of the photonic
subsystems. Local interaction of the photon-atom subsystems are of two-photon transition types and
thus we expect quadrature correlations directly interact with the atomic spin.

The quadrature operators for the field are defined as
\begin{subequations}\label{grp}
\begin{eqnarray}
X_{1}=\frac{1}{2}(a^{\dagger}+a),\\
X_{2}=\frac{i}{2}(a^{\dagger}-a).
\end{eqnarray}
\end{subequations}\label{grp}
Using the reduced density operators acting on the photonic subsystems, it can be readily shown that:
\begin{subequations}\label{grp}
\begin{eqnarray}
  (\Delta X_{1})^2=(\Delta X_{2})^2&=&\frac{1}{4}+\frac{1}{2}\lvert A\lvert^2,\\
  \nonumber (\Delta X_{1})^2=(\Delta X_{2})^2&=&\frac{7}{8}\lvert A\lvert^2+\frac{1}{4}(AC^{*}+CA^{*}+\lvert D\lvert^2)\\
  &+&\frac{1}{2}\lvert B\lvert^2,
\end{eqnarray}
\end{subequations}\label{grp}
\begin{figure}[!ht]
\includegraphics[width=8cm]{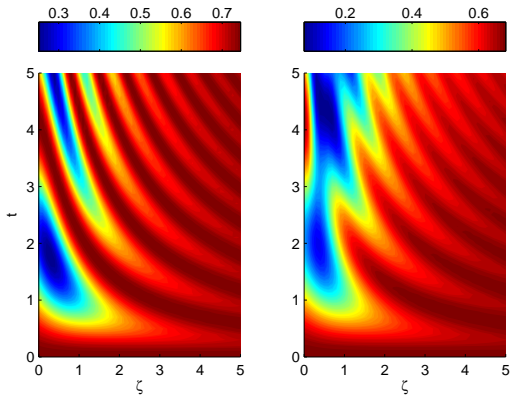}
\caption{(Color Online)
Variation of the variances of the quadratures for (left) initially entangled state and (right) initially non-entangled state with respect to time and hopping constant. Quadratures are squeezed for values $\leq 0.25$. At $t=0$, $(\Delta X_{1})^2=(\Delta X_{2})^2=0.75$ for initially entangled state and $(\Delta X_{1})^2=(\Delta X_{2})^2=0.6875$ for initially non-entangled state. All the parameters plotted are dimensionless as explained in the text.}
\label{fig:fig3}
\end{figure}
for initially entangled and initially non-entangled states respectively. As we see form the Fig.\ref{fig:fig3}, for initially entangled state (left) we have ideally squeezed states at darker regions where the product of the variances of the quadratures is equal to $0.25$ (since $(\Delta X_{1})^2=(\Delta X_{2})^2$). The behaviour of the quadratures is very similar to that of spin squeezing inequalities for initially entangled field (see Fig.\ref{fig:fig1}).

In the case that the state is initially non-entangled, situation is even more revealing. We see from the Fig. \ref{fig:fig3} (right) that initially there is no quadrature squeezing. But at later times we observe such correlations; especially at the darker regions we have $(\Delta X_{1})^2=(\Delta X_{2})^2\leq 1/4$.
Thus, in the light of above observations about quadratures, we conclude that (i) photon subsystem corresponds to effectively a spin system that is, under two-axis twisting interaction, squeezed, and particle entangled (ii) atoms do not directly interact with this effective spin system, but couple to the field by
two-photon transitions that can be linked to quadrature squeezing. (iii) Sharing the quadrature squeezed
particle entangled photons, the atoms build up sufficient noise reduction in particular directions to become spin squeezed and hence get particle entangled.
%
\section{conclusion} \label{sec:conclusion}
Summarizing, we examine a system of two cavities containing single atoms coupled to the cavity
field by two-photon transitions. The cavities are coupled by two-photon hopping interaction. We write
the coupling as two-axis twisting model of spin squeezing, which is also equivalent to LMG model.
We investigated spin squeezing and particle entanglement in atom and photon subsystems.
We consider two cases: one of them is initially entangled and the other is initially non-entangled. In both cases, we analytically and numerically determine that both spin squeezing and entanglement are
transferred from the photonic subsystem to the atomic subsystem and vice versa.
We briefly argued that coupling of the atomic collective spin variable and the effective photonic spin
is not a spin-spin interaction but of the form in which the quadrature correlations seem to be playing
a crucial role. We numerically verified that the transfer of spin squeezing between photons and atoms could be mediated by quadrature squeezing.
\acknowledgements
We gratefully acknowledge inspiring comments by J. Vidal.
\"O. E. M. acknowledges support by National Science and Technology Foundation of Turkey (T\"UB\.{I}TAK) under Project No. 109T267 and Project No. 111T285.

\bibliographystyle{apsrev}
\bibliography{squeezing_transfer.bib}

\begin{thebibliography}{32}
\expandafter\ifx\csname natexlab\endcsname\relax\def\natexlab#1{#1}\fi
\expandafter\ifx\csname bibnamefont\endcsname\relax
  \def\bibnamefont#1{#1}\fi
\expandafter\ifx\csname bibfnamefont\endcsname\relax
  \def\bibfnamefont#1{#1}\fi
\expandafter\ifx\csname citenamefont\endcsname\relax
  \def\citenamefont#1{#1}\fi
\expandafter\ifx\csname url\endcsname\relax
  \def\url#1{\texttt{#1}}\fi
\expandafter\ifx\csname urlprefix\endcsname\relax\def\urlprefix{URL }\fi
\providecommand{\bibinfo}[2]{#2}
\providecommand{\eprint}[2][]{\url{#2}}

\bibitem[{\citenamefont{Alexanian}(2011)}]{PhysRevA.83.023814}
\bibinfo{author}{\bibfnamefont{M.}~\bibnamefont{Alexanian}},
  \bibinfo{journal}{Phys. Rev. A} \textbf{\bibinfo{volume}{83}},
  \bibinfo{pages}{023814} (\bibinfo{year}{2011}).

\bibitem[{\citenamefont{Longhi}(2011)}]{longhi2011optical}
\bibinfo{author}{\bibfnamefont{S.}~\bibnamefont{Longhi}},
  \bibinfo{journal}{Physical Review A} \textbf{\bibinfo{volume}{83}},
  \bibinfo{pages}{43835} (\bibinfo{year}{2011}).

\bibitem[{\citenamefont{Zhou et~al.}(2009)\citenamefont{Zhou, Zhang, and
  Guo}}]{PhysRevA.80.013605}
\bibinfo{author}{\bibfnamefont{X.-F.} \bibnamefont{Zhou}},
  \bibinfo{author}{\bibfnamefont{Y.-S.} \bibnamefont{Zhang}}, \bibnamefont{and}
  \bibinfo{author}{\bibfnamefont{G.-C.} \bibnamefont{Guo}},
  \bibinfo{journal}{Phys. Rev. A} \textbf{\bibinfo{volume}{80}},
  \bibinfo{pages}{013605} (\bibinfo{year}{2009}).

\bibitem[{\citenamefont{Kitagawa and Ueda}(1993)}]{PhysRevA.47.5138}
\bibinfo{author}{\bibfnamefont{M.}~\bibnamefont{Kitagawa}} \bibnamefont{and}
  \bibinfo{author}{\bibfnamefont{M.}~\bibnamefont{Ueda}},
  \bibinfo{journal}{Phys. Rev. A} \textbf{\bibinfo{volume}{47}},
  \bibinfo{pages}{5138} (\bibinfo{year}{1993}).

\bibitem[{\citenamefont{S\o{}rensen et~al.}(2001)\citenamefont{S\o{}rensen,
  Duan, Cirac, and Zoller}}]{sorensen2001many}
\bibinfo{author}{\bibfnamefont{A.}~\bibnamefont{S\o{}rensen}},
  \bibinfo{author}{\bibfnamefont{L.~M.} \bibnamefont{Duan}},
  \bibinfo{author}{\bibfnamefont{J.~I.} \bibnamefont{Cirac}}, \bibnamefont{and}
  \bibinfo{author}{\bibfnamefont{P.}~\bibnamefont{Zoller}},
  \bibinfo{journal}{Nature} \textbf{\bibinfo{volume}{409}}, \bibinfo{pages}{63}
  (\bibinfo{year}{2001}).

\bibitem[{\citenamefont{Hillery}(2009)}]{hillery2009introduction}
\bibinfo{author}{\bibfnamefont{M.}~\bibnamefont{Hillery}},
  \bibinfo{journal}{Acta Physica Slovaca. Reviews and Tutorials}
  \textbf{\bibinfo{volume}{59}}, \bibinfo{pages}{1} (\bibinfo{year}{2009}).

\bibitem[{\citenamefont{Dell’Anno et~al.}(2006)\citenamefont{Dell’Anno,
  De~Siena, and Illuminati}}]{dell2006multiphoton}
\bibinfo{author}{\bibfnamefont{F.}~\bibnamefont{Dell’Anno}},
  \bibinfo{author}{\bibfnamefont{S.}~\bibnamefont{De~Siena}}, \bibnamefont{and}
  \bibinfo{author}{\bibfnamefont{F.}~\bibnamefont{Illuminati}},
  \bibinfo{journal}{Physics reports} \textbf{\bibinfo{volume}{428}},
  \bibinfo{pages}{53} (\bibinfo{year}{2006}).

\bibitem[{\citenamefont{Biswas and Agarwal}(2004)}]{PhysRevA.70.022323}
\bibinfo{author}{\bibfnamefont{A.}~\bibnamefont{Biswas}} \bibnamefont{and}
  \bibinfo{author}{\bibfnamefont{G.~S.} \bibnamefont{Agarwal}},
  \bibinfo{journal}{Phys. Rev. A} \textbf{\bibinfo{volume}{70}},
  \bibinfo{pages}{022323} (\bibinfo{year}{2004}).

\bibitem[{\citenamefont{Cirac et~al.}(1997)\citenamefont{Cirac, Zoller, Kimble,
  and Mabuchi}}]{PhysRevLett.78.3221}
\bibinfo{author}{\bibfnamefont{J.~I.} \bibnamefont{Cirac}},
  \bibinfo{author}{\bibfnamefont{P.}~\bibnamefont{Zoller}},
  \bibinfo{author}{\bibfnamefont{H.~J.} \bibnamefont{Kimble}},
  \bibnamefont{and} \bibinfo{author}{\bibfnamefont{H.}~\bibnamefont{Mabuchi}},
  \bibinfo{journal}{Phys. Rev. Lett.} \textbf{\bibinfo{volume}{78}},
  \bibinfo{pages}{3221} (\bibinfo{year}{1997}).

\bibitem[{\citenamefont{Vernac et~al.}(2001)\citenamefont{Vernac, Pinard, and
  Giacobino}}]{vernac2001quantum}
\bibinfo{author}{\bibfnamefont{L.}~\bibnamefont{Vernac}},
  \bibinfo{author}{\bibfnamefont{M.}~\bibnamefont{Pinard}}, \bibnamefont{and}
  \bibinfo{author}{\bibfnamefont{E.}~\bibnamefont{Giacobino}},
  \bibinfo{journal}{Eur. Phys. J. D} \textbf{\bibinfo{volume}{17}},
  \bibinfo{pages}{125} (\bibinfo{year}{2001}).

\bibitem[{\citenamefont{Hald et~al.}(1999)\citenamefont{Hald, S\o{}rensen,
  Schori, and Polzik}}]{PhysRevLett.83.1319}
\bibinfo{author}{\bibfnamefont{J.}~\bibnamefont{Hald}},
  \bibinfo{author}{\bibfnamefont{J.~L.} \bibnamefont{S\o{}rensen}},
  \bibinfo{author}{\bibfnamefont{C.}~\bibnamefont{Schori}}, \bibnamefont{and}
  \bibinfo{author}{\bibfnamefont{E.~S.} \bibnamefont{Polzik}},
  \bibinfo{journal}{Phys. Rev. Lett.} \textbf{\bibinfo{volume}{83}},
  \bibinfo{pages}{1319} (\bibinfo{year}{1999}).

\bibitem[{\citenamefont{Kuzmich et~al.}(1997)\citenamefont{Kuzmich, M\o{}lmer,
  and Polzik}}]{PhysRevLett.79.4782}
\bibinfo{author}{\bibfnamefont{A.}~\bibnamefont{Kuzmich}},
  \bibinfo{author}{\bibfnamefont{K.}~\bibnamefont{M\o{}lmer}},
  \bibnamefont{and} \bibinfo{author}{\bibfnamefont{E.~S.}
  \bibnamefont{Polzik}}, \bibinfo{journal}{Phys. Rev. Lett.}
  \textbf{\bibinfo{volume}{79}}, \bibinfo{pages}{4782} (\bibinfo{year}{1997}).

\bibitem[{\citenamefont{Banerjee}(1996)}]{PhysRevA.54.5327}
\bibinfo{author}{\bibfnamefont{A.}~\bibnamefont{Banerjee}},
  \bibinfo{journal}{Phys. Rev. A} \textbf{\bibinfo{volume}{54}},
  \bibinfo{pages}{5327} (\bibinfo{year}{1996}).

\bibitem[{\citenamefont{T\'oth et~al.}(2007)\citenamefont{T\'oth, Knapp,
  G\"uhne, and Briegel}}]{PhysRevLett.99.250405}
\bibinfo{author}{\bibfnamefont{G.}~\bibnamefont{T\'oth}},
  \bibinfo{author}{\bibfnamefont{C.}~\bibnamefont{Knapp}},
  \bibinfo{author}{\bibfnamefont{O.}~\bibnamefont{G\"uhne}}, \bibnamefont{and}
  \bibinfo{author}{\bibfnamefont{H.~J.} \bibnamefont{Briegel}},
  \bibinfo{journal}{Phys. Rev. Lett.} \textbf{\bibinfo{volume}{99}},
  \bibinfo{pages}{250405} (\bibinfo{year}{2007}).

\bibitem[{\citenamefont{Ma et~al.}(2011)\citenamefont{Ma, Wang, Sun, and
  Nori}}]{ma2011quantum}
\bibinfo{author}{\bibfnamefont{J.}~\bibnamefont{Ma}},
  \bibinfo{author}{\bibfnamefont{X.}~\bibnamefont{Wang}},
  \bibinfo{author}{\bibfnamefont{C.}~\bibnamefont{Sun}}, \bibnamefont{and}
  \bibinfo{author}{\bibfnamefont{F.}~\bibnamefont{Nori}},
  \bibinfo{journal}{Physics Reports}  (\bibinfo{year}{2011}).

\bibitem[{\citenamefont{Cunha et~al.}(2007)\citenamefont{Cunha, Dunningham, and
  Vedral}}]{cunha2007entanglement}
\bibinfo{author}{\bibfnamefont{M.}~\bibnamefont{Cunha}},
  \bibinfo{author}{\bibfnamefont{J.}~\bibnamefont{Dunningham}},
  \bibnamefont{and} \bibinfo{author}{\bibfnamefont{V.}~\bibnamefont{Vedral}},
  \bibinfo{journal}{Proceedings of the Royal Society A: Mathematical, Physical
  and Engineering Science} \textbf{\bibinfo{volume}{463}},
  \bibinfo{pages}{2277} (\bibinfo{year}{2007}).

\bibitem[{\citenamefont{Van~Enk}(2005)}]{van2005single}
\bibinfo{author}{\bibfnamefont{S.}~\bibnamefont{Van~Enk}},
  \bibinfo{journal}{Physical Review A} \textbf{\bibinfo{volume}{72}},
  \bibinfo{pages}{64306} (\bibinfo{year}{2005}).

\bibitem[{\citenamefont{Benatti et~al.}(2011)\citenamefont{Benatti, Floreanini,
  and Marzolino}}]{benatti2011entanglement}
\bibinfo{author}{\bibfnamefont{F.}~\bibnamefont{Benatti}},
  \bibinfo{author}{\bibfnamefont{R.}~\bibnamefont{Floreanini}},
  \bibnamefont{and}
  \bibinfo{author}{\bibfnamefont{U.}~\bibnamefont{Marzolino}},
  \bibinfo{journal}{Journal of Physics B: Atomic, Molecular and Optical
  Physics} \textbf{\bibinfo{volume}{44}}, \bibinfo{pages}{091001}
  (\bibinfo{year}{2011}).

\bibitem[{\citenamefont{{\"O}ztop et~al.}(2009)\citenamefont{{\"O}ztop, Oktel,
  M{\"u}stecapl{\i}o{\u{g}}lu, and You}}]{oztop2009quantum}
\bibinfo{author}{\bibfnamefont{B.}~\bibnamefont{{\"O}ztop}},
  \bibinfo{author}{\bibfnamefont{M.~{\"O}.} \bibnamefont{Oktel}},
  \bibinfo{author}{\bibfnamefont{{\"O}.~E.}
  \bibnamefont{M{\"u}stecapl{\i}o{\u{g}}lu}}, \bibnamefont{and}
  \bibinfo{author}{\bibfnamefont{L.}~\bibnamefont{You}}, \bibinfo{journal}{J.
  Phys. B} \textbf{\bibinfo{volume}{42}}, \bibinfo{pages}{145505}
  (\bibinfo{year}{2009}).

\bibitem[{\citenamefont{T{\'o}th et~al.}(2009)\citenamefont{T{\'o}th, Knapp,
  G{\"u}hne, and Briegel}}]{toth2009spin}
\bibinfo{author}{\bibfnamefont{G.}~\bibnamefont{T{\'o}th}},
  \bibinfo{author}{\bibfnamefont{C.}~\bibnamefont{Knapp}},
  \bibinfo{author}{\bibfnamefont{O.}~\bibnamefont{G{\"u}hne}},
  \bibnamefont{and} \bibinfo{author}{\bibfnamefont{H.}~\bibnamefont{Briegel}},
  \bibinfo{journal}{Physical Review A} \textbf{\bibinfo{volume}{79}},
  \bibinfo{pages}{042334} (\bibinfo{year}{2009}).

\bibitem[{\citenamefont{Vidal}(2006)}]{PhysRevA.73.062318}
\bibinfo{author}{\bibfnamefont{J.}~\bibnamefont{Vidal}},
  \bibinfo{journal}{Phys. Rev. A} \textbf{\bibinfo{volume}{73}},
  \bibinfo{pages}{062318} (\bibinfo{year}{2006}).

\bibitem[{\citenamefont{Vitagliano et~al.}(2011)\citenamefont{Vitagliano,
  Hyllus, Egusquiza, and T\'oth}}]{PhysRevLett.107.240502}
\bibinfo{author}{\bibfnamefont{G.}~\bibnamefont{Vitagliano}},
  \bibinfo{author}{\bibfnamefont{P.}~\bibnamefont{Hyllus}},
  \bibinfo{author}{\bibfnamefont{I.~n.~L.} \bibnamefont{Egusquiza}},
  \bibnamefont{and} \bibinfo{author}{\bibfnamefont{G.}~\bibnamefont{T\'oth}},
  \bibinfo{journal}{Phys. Rev. Lett.} \textbf{\bibinfo{volume}{107}},
  \bibinfo{pages}{240502} (\bibinfo{year}{2011}).

\bibitem[{\citenamefont{Wu and Yang}(1997)}]{wu1997effective}
\bibinfo{author}{\bibfnamefont{Y.}~\bibnamefont{Wu}} \bibnamefont{and}
  \bibinfo{author}{\bibfnamefont{X.}~\bibnamefont{Yang}},
  \bibinfo{journal}{Physical Review A} \textbf{\bibinfo{volume}{56}},
  \bibinfo{pages}{2443} (\bibinfo{year}{1997}).

\bibitem[{\citenamefont{Wu}(1996)}]{wu1996effective}
\bibinfo{author}{\bibfnamefont{Y.}~\bibnamefont{Wu}},
  \bibinfo{journal}{Physical Review A} \textbf{\bibinfo{volume}{54}},
  \bibinfo{pages}{1586} (\bibinfo{year}{1996}).

\bibitem[{\citenamefont{Alexanian and Bose}(1995)}]{PhysRevA.52.2218}
\bibinfo{author}{\bibfnamefont{M.}~\bibnamefont{Alexanian}} \bibnamefont{and}
  \bibinfo{author}{\bibfnamefont{S.~K.} \bibnamefont{Bose}},
  \bibinfo{journal}{Phys. Rev. A} \textbf{\bibinfo{volume}{52}},
  \bibinfo{pages}{2218} (\bibinfo{year}{1995}).

\bibitem[{\citenamefont{Alexanian et~al.}(1998)\citenamefont{Alexanian, Bose,
  and Chow}}]{alexanian1998trapping}
\bibinfo{author}{\bibfnamefont{M.}~\bibnamefont{Alexanian}},
  \bibinfo{author}{\bibfnamefont{S.}~\bibnamefont{Bose}}, \bibnamefont{and}
  \bibinfo{author}{\bibfnamefont{L.}~\bibnamefont{Chow}}, \bibinfo{journal}{J.
  Mod. Opt.} \textbf{\bibinfo{volume}{45}}, \bibinfo{pages}{2519}
  (\bibinfo{year}{1998}).

\bibitem[{\citenamefont{Vidal et~al.}(2004)\citenamefont{Vidal, Palacios, and
  Aslangul}}]{PhysRevA.70.062304}
\bibinfo{author}{\bibfnamefont{J.}~\bibnamefont{Vidal}},
  \bibinfo{author}{\bibfnamefont{G.}~\bibnamefont{Palacios}}, \bibnamefont{and}
  \bibinfo{author}{\bibfnamefont{C.}~\bibnamefont{Aslangul}},
  \bibinfo{journal}{Phys. Rev. A} \textbf{\bibinfo{volume}{70}},
  \bibinfo{pages}{062304} (\bibinfo{year}{2004}).

\bibitem[{\citenamefont{Ribeiro et~al.}(2008)\citenamefont{Ribeiro, Vidal, and
  Mosseri}}]{PhysRevE.78.021106}
\bibinfo{author}{\bibfnamefont{P.}~\bibnamefont{Ribeiro}},
  \bibinfo{author}{\bibfnamefont{J.}~\bibnamefont{Vidal}}, \bibnamefont{and}
  \bibinfo{author}{\bibfnamefont{R.}~\bibnamefont{Mosseri}},
  \bibinfo{journal}{Phys. Rev. E} \textbf{\bibinfo{volume}{78}},
  \bibinfo{pages}{021106} (\bibinfo{year}{2008}).

\bibitem[{\citenamefont{Ribeiro et~al.}(2007)\citenamefont{Ribeiro, Vidal, and
  Mosseri}}]{PhysRevLett.99.050402}
\bibinfo{author}{\bibfnamefont{P.}~\bibnamefont{Ribeiro}},
  \bibinfo{author}{\bibfnamefont{J.}~\bibnamefont{Vidal}}, \bibnamefont{and}
  \bibinfo{author}{\bibfnamefont{R.}~\bibnamefont{Mosseri}},
  \bibinfo{journal}{Phys. Rev. Lett.} \textbf{\bibinfo{volume}{99}},
  \bibinfo{pages}{050402} (\bibinfo{year}{2007}).

\bibitem[{\citenamefont{Karassiov and Klimov}(1994)}]{karassiov1994}
\bibinfo{author}{\bibfnamefont{V.}~\bibnamefont{Karassiov}} \bibnamefont{and}
  \bibinfo{author}{\bibfnamefont{A.}~\bibnamefont{Klimov}},
  \bibinfo{journal}{Physics Letters A} \textbf{\bibinfo{volume}{189}},
  \bibinfo{pages}{43} (\bibinfo{year}{1994}).

\bibitem[{\citenamefont{Higgs}(1979)}]{higgs1979}
\bibinfo{author}{\bibfnamefont{P.}~\bibnamefont{Higgs}},
  \bibinfo{journal}{Journal of Physics A: Mathematical and General}
  \textbf{\bibinfo{volume}{12}}, \bibinfo{pages}{309} (\bibinfo{year}{1979}).

\bibitem[{\citenamefont{Carmichael}(1986)}]{carmichael1986}
\bibinfo{author}{\bibfnamefont{H.~J.} \bibnamefont{Carmichael}},
  \bibinfo{journal}{Phys. Rev. A} \textbf{\bibinfo{volume}{33}},
  \bibinfo{pages}{3262} (\bibinfo{year}{1986}).

\end{thebibliography}
\end{document}